\PassOptionsToPackage{table,xcdraw}{xcolor}

\documentclass[sigconf]{acmart}

\AtBeginDocument{%
  \providecommand\BibTeX{{%
    \normalfont B\kern-0.5em{\scshape i\kern-0.25em b}\kern-0.8em\TeX}}}

\setcopyright{none}
\copyrightyear{2025}
\acmYear{2025}

\acmConference[ASSETS '25 Workshop: AT @ Work]{}{October 2025}{Virtual}

\usepackage{xpatch}

\makeatletter
\xpatchcmd{\ps@firstpagestyle}{Manuscript submitted to ACM}{}{\typeout{First patch succeeded}}{\typeout{first patch failed}}
\xpatchcmd{\ps@standardpagestyle}{Manuscript submitted to ACM}{}{\typeout{Second patch succeeded}}{\typeout{Second patch failed}}    \@ACM@manuscriptfalse
\makeatother

\acmBooktitle{Proceedings of AT @ Work: Intelligent Assistive Technologies for Enabling Workplace Inclusion at The 27th International ACM SIGACCESS
Conference on Computers and Accessibility (ASSETS ’25)} 
\acmISBN{}

\usepackage{appendix}
\usepackage{soul} 

\usepackage{subfigure}
\usepackage{subcaption}

\usepackage{graphicx}
\usepackage{multirow}
\usepackage{pifont}
\usepackage{lscape}

\usepackage{mathtools}

\begin{document}

\title[Creation, Critique, and Consumption: Exploring Generative AI Descriptions for Supporting \\Blind and Low Vision Professionals with Visual Tasks]{Creation, Critique, and Consumption: Exploring Generative AI Descriptions for Supporting Blind and Low Vision Professionals with Visual Tasks}

\author{Lucy Jiang}
\affiliation{%
  \institution{University of Washington}
  \city{Seattle, WA}
  \country{USA}}
\email{lucjia@uw.edu}

\author{Lotus Zhang}
\affiliation{%
  \institution{University of Washington}
  \city{Seattle, WA}
  \country{USA}}
\email{hanziz@uw.edu}

\author{Leah Findlater}
\affiliation{%
  \institution{University of Washington}
  \city{Seattle, WA}
  \country{USA}}
\email{leahkf@uw.edu}


\begin{abstract}
Many blind and low vision (BLV) people are excluded from professional roles that may involve visual tasks due to access barriers and persisting stigmas. Advancing generative AI systems can support BLV people through providing contextual and personalized visual descriptions for creation, critique, and consumption. In this workshop paper, we provide design suggestions for how visual descriptions can be better contextualized for multiple professional tasks. We conclude by discussing how these designs can improve autonomy, inclusion, and skill development over time.
\end{abstract}






\maketitle
\section{Introduction}
The employment rate of disabled people is roughly one-third that of non-disabled people \cite{bureaulaborservices}. For blind and low vision (BLV) individuals, this disparity can be partially attributed to professional expectations for visual communication proficiency in terms of creation, critique, or consumption \cite{branham2015invisible}. For example, common tasks include creating slide decks, formatting documents, taking or posting photos, watching video training content, and providing feedback to a colleague on their visual output. While helpful, traditional accommodations are often limited to remediating a lack of access to basic visual information (e.g., \cite{stangl2020person, ahmetovic2021touch}) rather than enabling full participation in the workplace. As a result, this cycle of inaccessibility can create a critical barrier to BLV people’s success in professional spaces.

The emergence of advancing generative AI-powered visual description technologies presents a novel opportunity to provide BLV professionals with both contextual and personalized information tailored to specific visual communication tasks. For example, Chang et al. \cite{chang2024editscribe} developed a system to enable BLV people to nonvisually edit images using natural language loops and Zhang et al. \cite{zhang2024designing} explored BLV people’s thoughts on how AI could support their visual privacy management across personal and professional domains. 

Moving forward, we encourage researchers and practitioners to continue developing contextual and personalized visual description systems with the option of providing specialized insights about design principles, spatial layouts, and other aesthetic qualities that can be crucial for professional visual communication tasks. This workshop paper highlights two specific cases where AI can be used to help support BLV professionals in diverse roles: (1) filming and editing videos as an independent content creator and (2) creating marketing materials for a large advertising agency. We argue that designing AI description technologies that support BLV people’s visual communication proficiency can improve overall inclusion across a variety of teams, projects, and tasks.

\section{Design Suggestions for Visual Descriptions}
Blind and low vision people may engage in a variety of visual communication and presentation activities for employment purposes. Here, we describe and synthesize two scenarios in which visual descriptions can be highly specialized and personalized to better support BLV professionals.

\subsection{Filming and Editing Videos as an Independent Content Creator} 
Prior work has highlighted how BLV people often engage with video-based social media platforms to share their daily lives and engage in the creator economy \cite{zhang2023understanding, rong2022it, seo2017exploring}. However, these works also surfaced many access issues: BLV people had difficulties creating visually-engaging content \cite{zhang2023understanding}, they often could not access or verify the appearance of visual features such as filters \cite{rong2022it}, and it was difficult to keep up with trends when videos by other creators lacked descriptions \cite{jiang2024context, vandaele2024making}. Creators often felt that adhering to visually-oriented practices was required for professional success, regardless of whether that was determined by algorithms, peers, or potential employers \cite{zhang2024designing, rong2022it}. To address visual challenges associated with video editing, Huh et al. \cite{huh2023avscript} developed a text-based video editing system that integrated visual descriptions and speech. This enabled BLV people to more easily identify visual errors to edit out, but had some limitations with regard to fine-grained navigation.

To further support blind and low vision content creators in filming and editing videos, we recommend designing systems that integrate AI visual descriptions throughout the entire creative process. This can begin with generating descriptions focused on visual trend identification (e.g., describing common visual accompaniments, such as specific gestures or text on screen, to trending audio tracks). AI generated descriptions can also support BLV people in content creation through supporting the filming process, either by ensuring the subject is in frame in the desired position or by describing the captured content in detail to assist with composing a more artistic shot. Lastly, we suggest providing additional artistic information during the editing process beyond editing the video for content, ranging from descriptions of the video’s warmth to the accuracy of specific filters or special effects.

\subsection{Creating Marketing Materials for a Large Advertising Agency}
Working as a creative professional in a larger company can present unique challenges for BLV people, given that they must engage in collaborative workflows and produce content in multiple formats. Unlike independent content creation, working in larger companies requires rapid iteration on concepts, real-time collaboration, and adherence to strict brand guidelines and client specifications. Some prior work explored how to improve the accessibility of digital graphic creation. For example, when working on digital artboards, BLV people valued having visual feedback to ensure that their resulting boards were visually appealing to sighted or low vision collaborators \cite{zhang2023a11yboard}. Others have begun investigating how to support mixed-ability collaboration when using text editors and slide deck software (e.g., \cite{das2022co11ab, peng2022diffscriber}), but real-time collaboration in less structured visual environments remains a challenge.

To better support BLV professionals in advertising agency environments, we recommend tailoring AI-generated descriptions to support common visual communication tasks. To aid with either creation or critique, describing key content such as brand guidelines can be valuable for avoiding misunderstandings or unnecessary iterations. For example, if the employee is tasked to create an infographic for a company, precise color descriptions may be necessary to ensure that the colors accurately represent the brand. Furthermore, we suggest creating AI systems that can increase the accessibility of team collaboration. For visually-oriented collaboration activities, which may include wireframing or whiteboarding, we encourage future AI system designs to provide information at a big-picture (e.g., describing the overall visual appearance) as well as at the object (e.g., describing a cluster of sticky notes) level. We also recommend describing where each collaborator’s cursor is located as it may be a proxy of visual focus.

\section{Discussion and Conclusion}
As illuminated by prior work, major barriers to employment and upward mobility include education, mentorship, access, and peer perceptions \cite{afb, cha2024understanding}. In particular, BLV people face significant stigmas surrounding their degree of visual literacy (i.e., ability to interpret, understand, and evaluate visual information). To a degree, this may be as a result of ignorance: some may not understand that blindness is a spectrum and that many BLV people may still have some amount of sight or were not born without sight. However, in \textit{More than Meets the Eye: What Blindness Brings to Art} \cite{kleege2017more}, author Georgina Kleege argues that regardless of the onset or degree of vision loss, \textit{“the average totally, congenitally blind person knows infinitely more about what it means to be sighted than the average sighted person knows about what it means to be blind.”} Building on Kleege’s observation, BLV people’s lived experience in visually-oriented societies means that they often possess a deep understanding of visual metaphors, cultural associations, and aesthetic concepts. Future access technology should support and augment BLV people’s visual literacy rather than assume its absence.

Making visual communication tasks more accessible is a critical step towards increasing BLV representation in the workplace. As blind and low vision professionals showcase their creative capabilities within these roles, they pave the way for more inclusive design practices and workplace cultures that can benefit all employees. We believe that this can gradually reduce the stigma and improve BLV people’s autonomy, inclusion, and skill development over time.

\section{Workshop Involvement}
\subsection{Author Biographies}
Lucy Jiang is a PhD student in Human Centered Design and Engineering at the University of Washington. Her research focuses on making both digital content (e.g., videos, images) and physical spaces (e.g., public art) more accessible through multimodal and context-aware methods.

Lotus Zhang is a PhD candidate in Human Centered Design and Engineering at the University of Washington. She researches the accessibility of digital content creation by blind and low vision individuals across professional and recreational contexts. In particular, her research explores the potential and risks of AI involvement in BLV individuals’ creation processes.

Leah Findlater is a Professor in Human Centered Design and Engineering at the University of Washington, where she directs the Inclusive Design Lab and is an Associate Director of CREATE (Center for Research and Education on Accessible Technology and Experiences). Her research is in accessibility and human-centered machine learning.

\subsection{Statement of Interest in Workshop}
This workshop paper addresses how emerging generative AI technologies can be designed for multiple professions to support greater involvement of BLV people in the workplace. We are interested in participating in this workshop given our prior experience in researching art accessibility, BLV content creators, and disabled people’s involvement in creative pursuits. We are also excited to learn from other accessibility researchers about different applications of accessible technology in the workplace, especially across different countries and contexts.

\begin{acks}
This work was supported by Apple Inc. Leah Findlater is also employed by and has a conflict of interest with Apple Inc. Any views, opinions, findings, and conclusions or recommendations expressed in this material are those of the author(s) and should not be interpreted as reflecting the views, policies, or position, either expressed or implied, of Apple Inc.
\end{acks}

\bibliographystyle{ACM-Reference-Format}
\bibliography{references}

\end{document}